\documentstyle[12pt]{article}
\newcommand{\newsection}[1]{
\addtocounter{section}{1}
\setcounter{equation}{0}
\setcounter{subsection}{0}
\addcontentsline{toc}{section}{\protect
\numberline{\arabic{section}}{{\rm #1}}}
\vglue .6cm
\pagebreak[3]
\noindent{\bf  \thesection. #1}\nopagebreak[4]\par\vskip .3cm}
\newcommand{\newsubsection}[1]{
\addtocounter{subsection}{1}
\addcontentsline{toc}{subsection}{\protect
\numberline{\arabic{section}.\arabic{subsection}}{#1}}
\vglue .4cm
\pagebreak[3]
\noindent{\it \thesubsection. #1}\nopagebreak[4]\par\vskip .3cm}

\renewcommand{\theequation}{\thesection.\arabic{equation}}

\newcommand{\ben}{\begin{enumerate}}
\newcommand{\een}{\end{enumerate}}
\newlength{\extraspace}
\newlength{\extraspaces}
\newcounter{dummy}
\newcommand{\bc}{\begin{center}}
\newcommand{\ec}{\end{center}}
\newcommand{\be}{\begin{equation}
\addtolength{\abovedisplayskip}{\extraspaces}
\addtolength{\belowdisplayskip}{\extraspaces}
\addtolength{\abovedisplayshortskip}{\extraspace}
\addtolength{\belowdisplayshortskip}{\extraspace}}
\newcommand{\ee}{\end{equation}}

\newcommand{\ba}{\begin{eqnarray}
\addtolength{\abovedisplayskip}{\extraspaces}
\addtolength{\belowdisplayskip}{\extraspaces}
\addtolength{\abovedisplayshortskip}{\extraspace}
\addtolength{\belowdisplayshortskip}{\extraspace}}
\newcommand{\ea}{\end{eqnarray}}

\newcommand{\ban}{\begin{eqnarray*}
\addtolength{\abovedisplayskip}{\extraspaces}
\addtolength{\belowdisplayskip}{\extraspaces}
\addtolength{\abovedisplayshortskip}{\extraspace}
\addtolength{\belowdisplayshortskip}{\extraspace}}
\newcommand{\ean}{\end{eqnarray*}}
\newcommand{\baa}{                         
\addtocounter{equation}{1}
\setcounter{dummy}{\value{equation}}
\setcounter{equation}{0}
\renewcommand{\theequation}{\thesection.\arabic{dummy}\alph{equation}}
\begin{eqnarray}
\addtolength{\abovedisplayskip}{\extraspaces}
\addtolength{\belowdisplayskip}{\extraspaces}
\addtolength{\abovedisplayshortskip}{\extraspace}
\addtolength{\belowdisplayshortskip}{\extraspace}}
\newcommand{\eaa}{                                       
\end{eqnarray}
\setcounter{equation}{\value{dummy}}
\renewcommand{\theequation}{\thesection.\arabic{equation}}}

\input epsf

\newcounter{fignum}
\newcounter{tabel}

\newcounter{tabnum}
\newcounter{xxx}
\newcommand{\bl}{\begin{list}{({\it\roman{xxx}})}{\usecounter{xxx}}}
\newcommand{\el}{\end{list}}
\newcommand{\ppt}[1]{{\partial \over \partial t}}            
\newcommand{\ppx}[1]{{\partial \over \partial x}}            
\newcommand{\pqt}[1]{{\partial^2 \over \partial t^2}}            
\newcommand{\pqx}[1]{{\partial^2  \over \partial x^2}}            
\def\<{\langle}
\def\>{\rangle}

\newfont{\gothic}{eufm10 scaled\magstep1}

\renewcommand{\hat}{\widehat}

\begin{document}
\begin{titlepage}
\begin{flushleft}
\today
\end{flushleft}
\begin{center}
{\LARGE \bf Free particle states from Geometry}
\end{center}
\vskip 2cm
\begin{center}
\mbox{F.M.C.\ Witte}\\
{
\it
Julius School of Physics and Astronomy, 
Utrecht University\\
Leuvenlaan 4, 3584 CE, Utrecht\\
Netherlands}
\end{center}
\vskip 1.5cm
\begin{abstract}
In this letter I discuss how conformal geometric algebra  models for Euclidean and Minkowksi spaces determine allowed quantum mechanical statespaces for free particles I explicitly treat $(1+1)$-dimensional spacetime and $2$-dimensional Euclidean space.
\end{abstract}
\end{titlepage}
\newpage
\newsection{Introduction}
Studying spacetime from a quantum mechanical point of view raises tough questions about the nature of the elementary degrees of freedom that make up quantum spacetime \cite{qgrav}. Graphical metaphores have been used to describe what we might expect to find when we study
the structure of spacetime down to the Planck scale \cite{gravi}. Some suggest the events making up spacetime degenerate into a form of "dust" of disconnected events that only form something like a manifold in some average sense. Others prefer to picture quantum spacetime as a "foam" continuously tearing and reconnecting under the influence of local topological fluctuations. Loop quantum gravity has provided the quantisation of area and volume as major results, also pointing towards a sense of disconnectedness or discreteness at the deepest level. String theory indicates that quantum gravitational states atleast can be counted to give rise to well-defined notions of entropy in accord with semi-classical expectations. Still, the road to obtaining the right underlying degrees of freedom is hard, and far from complete. The study of the effect of a classical background geometry on the quantum properties of matter is just one of the areas
of interest. This letter deals with the simplest of such systems.

In recent years the study of the geometry of spacetime as well as euclidean spaces has received some new input from geometric algebra
\cite{hestenes}. This pairing of linear algebra with Clifford algebra generates a powerfull language to describe geometric structures in spacetime or, for example, three dimensional euclidean space. Subsequently it has also been applied to describe the geometry of curved space \cite{gagrav}. Recently geometric algebras generating a $5$-dimensional conformal model, $CGA_{5}$ for euclidean $3d$ space have been discussed by several authors \cite{gaconf}. These geometric algebras are defined on vectorspaces with an indefinite innerproduct. Typically the $n$-dimensional euclidean space is modelled through a $n+1$-dimensional Minkowksi space. The blades of the model geometric algebra describe not only points, lines and planes, but also point-pairs, circles and spheres and their intersections \cite{gaconf} . Points in space can, in this context, be viewed as so-called {\it dual spheres} of zero radius. The point at infinity playes a crucial role in these models, as it does in spacetime physics \cite{infty}. In \cite{witte1} this has been picked up and generalised for $d=3+1$, yielding a similar conformal model, $CGA_{7}$ for spacetime. Moreover, the symmetry group of the innerproduct in the modelspaces allows for a coordinate-free, versor description of the symmetry groups of the modelled spaces.  In this letter I discuss the implications of some of these results. For the sake of simplicity I restrict this discussion to two-dimensional settings.

\newsection{CGA models}
In this letter I will consider a quantum mechanical particle to be adequaltely described by a wavefunction of some sort that is part of a Hilbertspace of physical states. Suppose that the space in which this particle lives, the {\it target space}, is modelled by some higher-dimensional geometric algebra of a {\it model space}. Geometric structures in the target space can be modelled efficiently by blades. Let us consider an example, see \cite{gaconf} for details.

Two points $a$ and $b$ in a two-dimensional euclidean targetspace can be represented by two vectors ${\bf a}$ and ${\bf b}$ in a $(3+1)$ dimensional modelspace with a Minkowski metric. The set-up of these models is such that the squared euclidean distance, $d_{E}^{2}(a,b)$ is given by
\be
d_{E}^{2}(a,b) = {\bf a} \cdot {\bf b} \ ,
\ee
where the dot indicates the Minkowski inner-product. The two-dimensional euclidean plane is modelled as the section of the lightcone in the modelspace with the hyperplane of vectors satisfying
\be
{\bf x} \cdot {\bf \infty } = -1 \ .
\ee
Here ${\bf \infty}$ is an arbitrarilly chosen null-vector that models the {\it point at infinity} of the targetspace. The remaining base-vector is another arbitrary choice, ${\bf O}$, representing a choice of origin in the targetspace. Hence, the choice of basis in the modelspace amounts to choosing an origin and a point at infinity in the targetspace. This does not limit our discussion here in any way. The great utility of this set up reveals itself when discussing subspaces. In a two-dimensional context subspaces may appear to be rather dull, however the $CGA_{4}$ model of two-dimensional euclidean also allows for an easy definition of circles. The circle through the two points $a$ and $b$ in our example, and another point $c$ is specified by the blade
\be
S^{1} = {\bf a} \wedge {\bf b}\wedge {\bf c} \ ,
\ee
in the so-called direct representation as all the points $x$ for which
\be
{\bf x} \wedge S^{1} = 0 \ ,
\ee
are on the circle. We can also write down the dual representation. Here the dual of the blade $S^{1}$ is denoted $S^{1}_{D}$ and can be used to define the same circle as the set of all $x$ for which
\be
{\bf x} \cdot {S_{D}^{1}} = 0 \ .
\ee
This is even more usefull as the dual circle is represented by a vector in modelspace, and the innerproduct transforms trivially under the isometries of modelspace.

\newsection{Quantum States of Free particles}
Now that we have briefly discussed how conformal models work it is time to ask what relevance this has for the quantum states of free particles in the modelspaces. Again, I will argue along two-dimensional examples. I will resort to an approach similar to \cite{wigner} and that has also been used in \cite{spacetime} to analyse the dimensionality of spacetime.

\newsubsection{$2$-dimensional Euclidean targetspace}
The basis of modelspace typically consists of the vectors $\{ {\bf O}, {\bf \infty}, {\bf e}_{1} , {\bf e}_{2}\}$, where ${\bf e}_{i}^2 = 1$, and ${\bf e}_{1} \cdot {\bf e}_{2} = {\bf e}_{j} \cdot {\bf \infty} = {\bf e}_{j} \cdot {\bf O} = 0$. This is not a orthonormal set, but such a set is easilly constructed using
\be
{\bf e}_{\pm} = \frac{1}{\sqrt{2}} ({\bf O} \mp {\bf \infty}) \ .
\ee
We get for the squares of these vectors
\be
{\bf e}_{\pm}^2 = \pm 1 \ .
\ee
Now, the vector ${\bf e}_{+}$ represents a dual circle around the origin with unit radius. The full group of isometries of the modelspace in this case is $SO(3,1)$, i.e. the Lorentz-group. 

Although the modelspace vectors themselves are devoid of physical meaning, there are physical reasons to pick out one particular vector. The point at infinity, modelled by the vector ${\bf \infty}$, stands out. It usually is the point where boundary conditions on wavefunctions are implemented. Thus an isometry that moves this point about is definitely going to interfere with any physics in the modelspace. So we keep ${\bf \infty}$ fixed. The {\it little group} of a null-vector like ${\bf \infty}$ is the group of two-dimensional euclidean motions $E_{2}$ \cite{spacetime}. This is not very surprising, but consistent.

Suppose we would fix the wavefunction at a {\it circle} at infinity. This produces the more interesting instance of keeping a dual circle fixed. Infact, for the determination of the apropriate little group the size of the circle is immaterial. Hence, fixing ${\bf e}_{+}$ will do. The corresponding little group is $SO(2,1)$. So in this case we expect to find a space of physical states that is a unitary representation of $SO(2,1)$, which are labelled by two integers \cite{so21}. Here they acquire new meaning as quantum numbers labelling the physical states on a two-dimensional targetspace. Let us investigate the physical meaning of the generators of this little group. In term of modelspace vectors, they are
\ba
S_{1} & = & {\bf e}_{-}{\bf e}_{1} \nonumber  \ , \\
S_{2} & = &{\bf e}_{-}{\bf e}_{2} \ , \\
S_{3} & = & {\bf e}_{1}{\bf e}_{2} \ . \nonumber 
\ea
It is easy to check these generators indeed generate an $SO(2,1)$ algebra. What is more important, $S_{3}$ is the generator of rotations in the $2$-dimensional euclidean plane. Hence, if interpreted as a quantum mechanical operator its eigenvalues $m$,
\be
{\hat{S}}_{3} | k m \rangle = m | k m \rangle \ ,
\ee
would correspond to angular momentum. However, $SO(2,1)$ does not have finite dimensional unitary representations. So, $m$ is not bound. Infact, from all unitary representations in which $S_{3}$ is diagonalised, there are only two which will allow $m$ to assume both positive as well as negative values, which would seem physically reasonable. The quantum number $k$ is related to the eigenvalues of the Casimir operator
\be
{\hat{S}}^2 = {\hat{S}}_{3}^2 - {\hat{S}}_{1}^2  - {\hat{S}}_{2}^2 \ , 
\ee
through
\be
{\hat{S}}^2 | k m \rangle = k(k+1) | k m \rangle \ .
\ee
The operators $S_{1}$ and $S_{2}$ are related to the generators of translations, but not identical to these. The physical meaning of the Casimir operator therefor needs further clarification. More details on the $SO(2,1)$ representations can be found in \cite{so21}.

\newsubsection{$1+1$-dimensional target-Spacetime}
Now let us consider the same in two-dimensional Minkowski space. Writing down a $CGA_{4}$ model for $(1+1)$-dimensional spacetime requires a little more care with respect to treating infinity, as is discussed in \cite{witte1}. Yet such a model can be found. The full group of isometries in this case is $SO(2,2)$. If we fix the point at infinity we obtain a little group which is the semi-direct product of two translations and 
$SO(1,1)$. This group does not yield any new phenomenology.

In the $CGA_{4}$ models for spacetime we do get a new ingredient. Now non-null vectors in modelspace represent dual $1$-shells at a constant proper distance from a given event in spacetime. They can either be spacelike shells or time-like shells. Such spacelike shells occur as bubblewalls in bounce solutions for decaying metastable states. The timelike shells occur for example as hadronization surfaces in thedescription of heavy-ion collissions. If we fix a $2$-shell the corresponding little group is $SO(2,1)$ as above. Here the situation is more complicated as the generators of the corresponding $SO(2,1)$ little groups do not contain evidently physical operators. One of the operators generates the $(1+1)$-dimensional Lorentz transformations and does not correspond to physical quantities. The other two are related to time- or space-translations. For example,
\be
A_{0} = \frac{1}{\sqrt{2}} {\bf e}_{0} ({\bf O} - {\bf \infty})  \ ,
\ee
contains a part, ${\bf e}_{0} {\bf \infty}$, that generates time-translations and thus would relate to energy-eigenvalues. The other part can be dubbed {\it a timelike tangentvector at the origin} following Dorst \cite{gaconf}. A final interesting point to note here is that the little groups of timelike and spacelike $1$-shells coincide. This however is restricted to the $(1+1)$-dimensional case. In higher-dimensional target-spacetime the little groups will differ!

\newsection{Closing remarks}
I have made an analysis of quantum states on $2$-dimensional Euclidean and Minkowski targetspaces by means of a conformal geometric algebra modelspace. This yields interesting assertions about the spaces of physical quantumstates on these targetspaces. Let me close this letter by stating a few words about the generalisation to physical three and four dimensional situations.

If we consider $3$-dimensional euclidean targetspace, the modelspace will have a $SO(4,1)$ isometry group. The relevant non-trivial little
group will be $SO(3,1)$, which has a wellknown representation theory \cite{wigner}. For $3$-dimensional target-spacetime the modelspace has a $SO(3,2)$ isometry group. The non-trivial little groups should be $SO(3,1)$ and $SO(2,2)$ where the computation of the relevant unitary representations is known \cite{spacetime}. Timelike and spacelike shells have different little groups. For $4$-dimensional target-spacetime the work reported in \cite{witte1} suggests that the modelspace is at least $6$-dimensional with a $SO(4,2)$ isometry group, but possibly larger. The relevant non-trivial little groups are expected to have $SO(4,1)$ and $SO(3,2)$ subgroups, here the computations become much more involved, see \cite{so41} and \cite{so42}. The identification of physically relevant modelspace isometry generators is currently under investigation \cite{witte2}

\end{document}